\documentclass[useAMS,usenatbib]{mn2e}
\usepackage{epsfig}
\usepackage{float}

\usepackage{natbib}

\usepackage{color}

\usepackage{aas_macros}

\usepackage{array}
\newcolumntype{L}[1]{>{\raggedright\let\newline\\\arraybackslash\hspace{0pt}}m{#1}}
\newcolumntype{C}[1]{>{\centering\let\newline\\\arraybackslash\hspace{0pt}}m{#1}}
\newcolumntype{R}[1]{>{\raggedleft\let\newline\\\arraybackslash\hspace{0pt}}m{#1}}

\def\sige{\mbox{$\sigma_{\rm e}$}}

\def\Re{\mbox{$R_{\rm e}$}}

\def\Msun{\mbox{$M_\odot$}}

\def\ML{\mbox{$M/L$}}
\def\dimf{\mbox{$\delta_{\rm IMF}$}}

\def\Yssp{\mbox{$\Upsilon_{SSP}$}}
\def\Yst{\mbox{$\Upsilon_\star$}}

\def\Ydyn{\mbox{$\Upsilon_{\rm dyn}$}}
\def\Mdyn{\mbox{$M_{\rm dyn}$}}
\def\Ytot{\mbox{$\Upsilon_{\rm tot}$}}

\def\mst{\mbox{$M_{\star}$}}

\def\cMvir{\mbox{$c_{\rm vir}-M_{\rm vir}$}}

\def\Mvir{\mbox{$M_{\rm vir}$}}
\def\cvir{\mbox{$c_{\rm vir}$}}

\def\fdm{\mbox{$f_{\rm DM}$}}

\def\lsim{\mathrel{\rlap{\lower3.5pt\hbox{\hskip0.5pt$\sim$}}
    \raise0.5pt\hbox{$<$}}}                
\def\gsim{~\rlap{$>$}{\lower 1.0ex\hbox{$\sim$}}}

\def\sigAp{\mbox{$\sigma_{\rm Ap}$}}

\def\atlas3d{ATLAS$^{\rm 3D}$}

\def\a0{\mbox{$a_{\rm 0}$}}

\def\gN{\mbox{$g_{\rm N}$}}

\def\NFWf{\mbox{{\tt NFWf}}}
\def\NFWfmulti{\mbox{{\tt NFWf-multi}}}
\def\NFWfhc{\mbox{{\tt NFWf-hc}}}

\def\NFWfWMAPone{\mbox{{\tt NFWf-WMAP1}}}
\def\NFWfWMAPthree{\mbox{{\tt NFWf-WMAP3}}}

\def\NFWfmtani{\mbox{{\tt NFWf-mild-tan-$\beta$}}}
\def\NFWfstani{\mbox{{\tt NFWf-strong-tan-$\beta$}}}
\def\NFWfmrani{\mbox{{\tt NFWf-mild-rad-$\beta$}}}

\def\cMLf{\mbox{{\tt cMLf}}}
\def\MOND{\mbox{{\tt MOND}}}
\def\MONDone{\mbox{{\tt MOND1}}}
\def\MONDtwo{\mbox{{\tt MOND2}}}
\def\NFWC{\mbox{{\tt NFWC}}}
\def\cMLC{\mbox{{\tt cMLC}}}

\def\Fig{\mbox{Figure~}}

\def\Tab{\mbox{Table~}}
\def\Sec{\mbox{Section~}}

\def\Remaj{\mbox{$R_{\rm maj}$}}

\title[IMF and DM in dEs]{Dark Matter and IMF normalization in Virgo dwarf early-type galaxies}

\author[Tortora C. et al.]{\noindent
C.~Tortora$^{1}$\thanks{E-mail: ctortora@na.astro.it}, F.~La
Barbera$^{1}$, N.R.~Napolitano$^{1}$
\\~\\
$^1$ INAF -- Osservatorio Astronomico di Capodimonte, Salita
Moiariello, 16, 80131 - Napoli, Italy}

\begin{document}

\date{Accepted  Received }
\pagerange{\pageref{firstpage}--\pageref{lastpage}} \pubyear{xxxx}
\maketitle

\label{firstpage}
\begin{abstract}
In this work we analyze the dark matter (DM) fraction, \fdm, and
mass-to-light ratio mismatch parameter, \dimf\ (computed with
respect to a Milky-Way-like IMF), for a sample of 39 dwarf
early-type galaxies (dEs) in the Virgo cluster. Both \fdm\ and
\dimf\ are estimated within the central (one effective radius)
galaxy regions, with a Jeans dynamical analysis that relies on
galaxy velocity dispersions, structural parameters, and stellar
mass-to-light ratios from the SMAKCED survey. In this first
attempt to constrain, simultaneously, the IMF normalization and
the dark matter content, we explore the impact of different
assumptions on the DM model profile. On average, for a NFW
profile, the \dimf\ is consistent with a Chabrier-like
normalization ($\dimf \sim 1$), with $\fdm \sim 0.35$. One of the
main results of the present work is that for at least a few
systems the \dimf\ is heavier than the Milky-Way-like value (i.e.
either top- or bottom-heavy). When introducing tangential
anisotropy, larger \dimf\ and smaller \fdm\ are derived. Adopting
a steeper concentration--mass relation than that from simulations,
we find lower \dimf\ ($\lsim 1$) and larger \fdm . A constant \ML\
profile with null \fdm\ gives the heaviest \dimf\ ($\sim 2$). In
the MONDian framework, we find consistent results to those for our
reference NFW model. If confirmed, the large scatter of \dimf\ for
dEs would provide (further) evidence for a non-universal IMF in
early-type systems. On average, our reference \fdm\ estimates are
consistent with those found for low-\sige\ ($\rm \sim 100 \, \rm
km s^{-1}$) early-type galaxies (ETGs). Furthermore, we find \fdm\
consistent with values from the SMAKCED survey, and find a
double-value behavior of \fdm\ with stellar mass, which mirrors
the trend of dynamical \ML\ and global star formation efficiency
(from abundance matching estimates) with mass.
\end{abstract}


\section{Introduction}\label{sec:intro}

Dark matter (DM) is a ubiquitous component of the universe and
dominates the mass density of virialized objects as galaxies and
clusters of galaxies. The current scenario of galaxy evolution
predicts that structures form bottom-up, i.e. the smallest haloes
form first, and then, larger and more massive haloes are created
from the merging of such smaller objects. Within this scenario,
numerical simulations of (DM only) structure formation within the
standard $\Lambda$CDM framework have provided accurate predictions
on the DM density distribution from dwarf to massive galaxies, up
to bigger structures, such as clusters of galaxies
(\citealt{NFW96}; \citealt{Bullock+01}; \citealt{Maccio+08}).
However, more realistic models have been produced to evaluate the
effect of baryons on the DM distribution (e.g.,
\citealt{Gnedin+04}).

No model of galaxy formation can be complete without an
understanding of how dwarf galaxies form, as these systems are the
closest objects, in the nearby Universe, to the building blocks of present, more massive,
galaxies. Their shallow potential wells make them susceptible to a
large variety of processes, from supernova feedback, to externally
induced effects, such as photoionization and heating from cosmic
UV background, as well as environmental processes, such as tidal
interactions and ram-pressure stripping (\citealt{dek_birn06};
\citealt{Recchi14}). This makes dwarf galaxies challenging to
model, but at the same time, excellent laboratories to test
important ingredients of astrophysics.

Providing a picture of the formation and evolution of dwarf
galaxies implies to understand the origin of their luminous and
dark mass components. In this regard, together with age and
metallicity, the stellar Initial Mass Function (IMF) is a key
stellar ingredient, as varying the IMF can lead to variations of a
factor of $\sim 2$ into the mass scale of galaxies. Direct counts
in the Milky Way (MW) have originally characterized the IMF as a
power-law mass-distribution, $dN/dM \propto M^{-\alpha}$, with
$\alpha \sim 2.35$ \citep{Salpeter55}, and subsequently refined it
to flatten at lower masses ($M \lsim 0.5 M_\odot$;
\citealt{Kroupa01,Chabrier03}). The IMF has been initially
considered as universal across galaxy types and cosmic time,
mostly because of a lack of evidence of IMF variations among
stellar clusters and OB associations in our Galaxy. This
assumption has been recently questioned by a loud chorus of
dynamical, lensing, and stellar population studies, finding
evidence for systematic IMF variations in massive early-type
galaxies (ETGs; \citealt{Treu+10}; \citealt{ThomasJ+11};
\citealt{Conroy_vanDokkum12b}; \citealt{Cappellari+12,
Cappellari+13_ATLAS3D_XX}; \citealt{Spiniello+12};
\citealt{Wegner+12}; \citealt{Dutton+13}; \citealt{Ferreras+13};
\citealt{Goudfrooij_Kruijssen13};
\citealt{LaBarbera+13_SPIDERVIII_IMF}; \citealt{TRN13_SPIDER_IMF};
\citealt{Weidner+13_giant_ell}; \citealt{Goudfrooij_Kruijssen14};
\citealt{Shu+15_SLACSXII};
\citealt{Tortora+14_DMslope,Tortora+14_MOND}). These independent
lines of evidence are interpreted with an excess of low-mass stars
in high- (relative to low-)mass ETGs, implying lower DM fractions
in these systems than found under the assumption of a universal,
MW-like,  distribution (e.g., \citealt{Cappellari+13_ATLAS3D_XX};
\citealt{TRN13_SPIDER_IMF}). However, evidence for a heavier IMF
normalization, than the MW-like one, has been recently questioned
for three nearby massive galaxies by \cite{Smith14} and
\cite{SLC15}, based on a lensing analysis of low-redshift ETGs.

Previous works of the DM and stellar mass distribution in galaxies
have focused on the study of intermediate-luminosity and bright
ETGs, with stellar masses $\mst \gsim 10^{10}\, \rm \Msun$. At
lower mass scales, most of the analysis have assumed universal
IMF: e.g. \cite{Geha+02}, who have fitted long-slit spectroscopy
for six dwarf ellipticals (dE), \cite{Rys+14}, who have performed
the full dynamical modelling of 2D kinematic data of 12 dEs,
\cite{Toloba+14_II}, who use virial theorem to model effective
velocity dispersion for a sample of 39 dEs. In contrast, there are
very few analysis of the IMF, see, e.g., the direct constraints in
ultra-faint dwarf galaxies around the Milky Way
(\citealt{Geha+13}). Moreover, to date, no detailed dynamical
analysis has been performed to characterize both the DM content
and IMF normalization (i.e. the stellar mass-to-light ratio
normalized to that expected for a MW-like distribution) at these
low masses, e.g. in dEs. As more massive ETGs, dEs are an ideal
target for this kind of study, as they are dominated by old halo
stars, have negligible star formation rate at present, and little
dust content (\citealt{deLooze+10_HVCS_VII};
\citealt{Toloba+12_II}; \citealt{Toloba+14_II}), making the
computation of mass-to-light ratios less affected by systematics.

In this paper, we fill the above-mentioned gap, studying the
central mass distribution of 39 Virgo dEs, drawn from
\citet[hereafter T+14]{Toloba+14_II}, in the stellar mass range
$\sim 10^{8}-10^{9} \, \rm \Msun$. We fit the central dynamics
using various (stellar and DM) mass distribution profiles,
comparing our findings with results for massive systems. The
layout of the paper is the following. In \Sec\ref{sec:data} we
describe the datasample and our dynamical method. In
\Sec\ref{sec:results} we discuss the results of the paper.
Conclusions are drawn in \Sec\ref{sec:conclusions}.

\section{Samples and data analysis}\label{sec:data}

\subsection{Dataset}\label{subsec:data}

We analyze a sample of 39 dEs in the magnitude range $-19 < \rm
M_{\rm r} < -16$, selected from the Virgo Cluster Catalog (VCC,
\citealt{Binggeli+85}) to have high surface brightness with
respect to the dwarf galaxy population (\citealt{Janz_Lisker09}).
Albeit incomplete in luminosity, this sample is representative of
the early-type population in this magnitude range (T+14).

The analysis relies on the following data:
\begin{itemize}
\item  The H-band structural parameters (the major-axis effective radius, \Remaj, S\'ersic index, n,
and axis ratio, q) are taken from T+14 and \cite{Janz+14}. The
nucleus, if present, is excluded from the fit. As noted by
\cite{Janz+14}, while the global profiles of most galaxies are
well approximated by a single S\'ersic profile, a more detailed
inspection reveals some significant deviations from a single
component. For this reason, to analyze systematics in the modeling
of the light profile, we have also investigated the impact of both
single- and double-S\'ersic fits on our results (see
\Sec\ref{sec:results}).
\begin{itemize}
\item {\it Single S\'ersic fit.} The structural parameters are taken from
Table~4 of T+14, where the \Re\ from the growth curve fit and the
S\'ersic index from the single S\'ersic fit are from
\cite{Janz+14}. The \Re's of the growth curve fits match
reasonably well those from direct single-S\'ersic fits. For 9
systems without a measured value of $n$ (as they had no fit with a
single S\'ersic component or are not present in
\citealt{Janz+14}), we have adopted $n=1$, testing the impact for
different choices of this parameter (see \Sec\ref{sec:results}).
Although only 8 out of the 39 SMAKCED dEs are best fitted with a
single (relative to a double) S\'ersic profile (see below), to
allow a more homogeneous comparison with massive ETGs from the
SPIDER sample (\citealt{SPIDER-I}), we adopt single-component
parameters as our reference case throughout the present work.
\item {\it Double S\'ersic fit.}  Data for the inner and outer S\'ersic components are taken
from Table 5 in \cite{Janz+14}, for the 33 (out of SMAKCED sample
of 39) dEs analyzed in that work. Out of these 33 dEs, 25 objects
are better described by multiple components, with 19 galaxies
being described by a double S\'ersic fit and 6 by a single
S\'ersic profile plus a lens.
\end{itemize}

\item Effective velocity dispersions, \sige, computed within an ellipse
of semi-major axis length of one \Remaj\ (T+14). The \sige's have
been computed by~T+14 by flux-averaging both rotation velocity and
velocity dispersion within each galaxy isophote, hence accounting
for both ordered and random motions in each system.
\item Age and metallicity estimates are taken from Table~5 of T+14, which have fitted relevant Lick spectral
indices -- measured within the \Remaj\ ellipse -- with
\cite{Vazdekis+12} simple stellar population (SSP) models. Using
exponentially declining star formation histories, T+14 have
demonstrated that the stellar masses are, on average, fairly
consistent with the SSP estimates, but the scatter is larger. For
four galaxies (VCC 0170, VCC 0781, VCC 1304 and VCC 1684) $H\beta$
and/or $H\alpha$ are found in emission. The emission lines are
narrower than absorption lines, thus the two components can be
decoupled (see \citealt{Toloba+14_II} for details). Although they
do not find any significant emission in any of the other galaxies,
they cannot rule out the possibility of them having some emission.
Thus, for galaxies with undetected emission features the estimated
ages (and stellar masses) would be taken as upper limits. See
\Sec\ref{subsec:IMF_DM} for further details.
\end{itemize}

\subsection{Analysis}

For each galaxy, we obtain the stellar H-band mass-to-light (\ML)
ratio, \Yssp, using the best fitted age and metallicity from T+14,
and the simple stellar population (SSP) models
of~\cite{Vazdekis+12}, for a Kroupa IMF. These \Yssp\ are
converted to those for a Chabrier IMF, by subtracting $0.05$ dex
(i.e. the difference in normalization between the Kroupa and
Chabrier IMFs; \citealt{Tortora+09}). Under the assumption of a
radially constant \Yssp, the deprojected mass profile of the
stellar component is written as $\mst(r) = \Yst j_{*}(r)$, with
$\Yst = \Yssp$. To derive the light profile $j_{*}(r)$, we perform
a deprojection, under the assumption of spherical symmetry, of the
H-band S\'ersic profiles (from the galaxy structural parameters,
see above). Dynamical (DM + light) mass estimates are obtained by
fitting the observed \sige\ with spherical, isotropic, Jeans
equations (\citealt{Tortora+09}). To account for the fact that
\sige\ is averaged within an elliptic aperture, while we rely on
spherical models, we calculate the 3D velocity dispersion from the
radial Jeans equation at the circularized (geometric) effective
radius\footnote{This circularized quantity is used throughout this
work.} $\Re = \sqrt{q} \Remaj$. The dynamical (i.e. total) mass
distribution of galaxies is computed by adopting either
single-component (i.e. a radially constant dynamical \ML) or
two-component profiles (i.e. stellar component plus DM halo) with
the stellar \Yst\ being a free fitting parameter, or fixed to
\Yssp\ (in case some other quantity, e.g. concentration, is let
free to vary, see the different cases described below).

Thus, after the mass model is chosen and the predicted velocity
dispersion, $\sigma_{e}^{J}(p)$, from the Jeans equation, is
derived, the equation $\sigma_{e}^{J}(p) = \sige$ is solved with
respect to the free parameter p \footnote{The adopted galaxy
models and the parameter p are defined below, in
\Sec\ref{sec:models}.}. The uncertainties on the best-fitting
parameter p and derived quantities are obtained by shifting the
input parameters (i.e. \sige, \Re, n, \mst) according to their
errors  a number of times and considering the distributions of
corresponding best-fitting solutions.

\subsection{Mass models}\label{sec:models}

We rely only on velocity dispersions measured within a single
aperture, which does not allow us to constrain the shape of the DM
profile in detail.  To this effect, we explore a variety of
models, analyzing several plausible assumptions.

The range of models considered in this study are summarized in
Table~\ref{tab:tab1}. Numerical collisionless N\,-\,body
simulations have provided clues on the formation and the evolution
of DM haloes, finding that the DM density of the haloes (from
dwarf galaxies to clusters) is independent of halo mass and well
described by a double power\,-\,law relation with a cusp at the
center (\citealt{NFW96,NFW97}; \citealt{Moore+98}). Thus, it is a
natural choice to start from this theoretical motivated class of
DM profiles.

As reference case, we adopt the two-component mass profile \NFWf\
(see Table~\ref{tab:tab1}), given by a S\'ersic-based stellar mass
distribution (with $p \equiv \Yst$ as a free parameter) and a
standard DM halo with a \citet[NFW, hereafter]{NFW96} density
profile. We relate virial mass, $\Mvir$, to concentration,
$\cvir$, with the mean trend for a WMAP5 cosmology (for relaxed
halos in \citealt{Maccio+08}), while the \Mvir--\mst\ relation is
assumed from \citet[M+10 hereafter]{Moster+10}, which extends down
to stellar masses of $\sim 10^{8}\, \rm \Msun$.

The model \NFWfmulti\ is used to study the impact of varying the
parametrization of the galaxy light distribution. For galaxies
whose light distribution is better fitted by a multiple, rather
than a single, component according to \cite{Janz+14}, we describe
the stellar component with double-S\'ersic fit parameters.

Following \cite{Tortora+14_DMslope}, we also explore how our
results depend on the assumed \cvir -- \Mvir\ relation. In
particular, since for higher mass galaxies (than those analyzed in
the present work), some studies suggest higher concentrations than
those from simulations~\citep[hereafter LFS12]{Buote+07,
Leier+12}, we also consider ``high-concentration'' models
(\NFWfhc), with $\cvir = 20$, in contrast to the typical value of
$\sim 12$ predicted for our dEs from the \citep{Maccio+08}
relation (for $\mst \sim 10^{9} \, \rm \Msun$ and $\Mvir \sim
10^{11}\, \rm \Msun$). We refer to these models as
``high-concentration'' NFW models, \NFWfhc. Moreover, the impact
of cosmological framework on the theoretical \cMvir\ relation is
analyzed with models \NFWfWMAPone\ and \NFWfWMAPthree, which use
WMAP1 and WMAP3 results from N-body simulations in
\cite{Maccio+08}. Notice that WMAP1 predictions are very similar
to the \cMvir\ based on the first release of Planck cosmological
parameters (\citealt{Dutton_Maccio14}).

A possible source of systematics is the hypothesis of isotropic
stellar orbits, as spatially resolved stellar kinematics for a
handful of dEs has been found to be better modelled with
tangential anisotropies, rather than isotropy (e.g.
\citealt{Geha+02}). Although a detailed analysis of galaxy
anisotropies is far from being trivial, and is certainly beyond
the scope of the present work, we have estimated the impact of
anisotropy on our inferences. To this effect, in our list of
models (Table~\ref{tab:tab1}), we have also included three cases
corresponding to radially constant anisotropy $\beta$ in the Jeans
equations (see also \citealt{Tortora+09, SPIDER-VI}): a ``mild''
tangential anisotropy, $\beta=-0.4$, (\NFWfmtani), a ``strong''
tangential anisotropy, $\beta = -1$, (\NFWfstani) and a ``mild''
radial anisotropy, $\beta = 0.4$ (\NFWfmrani).

\begin{figure*}
\centering \psfig{file=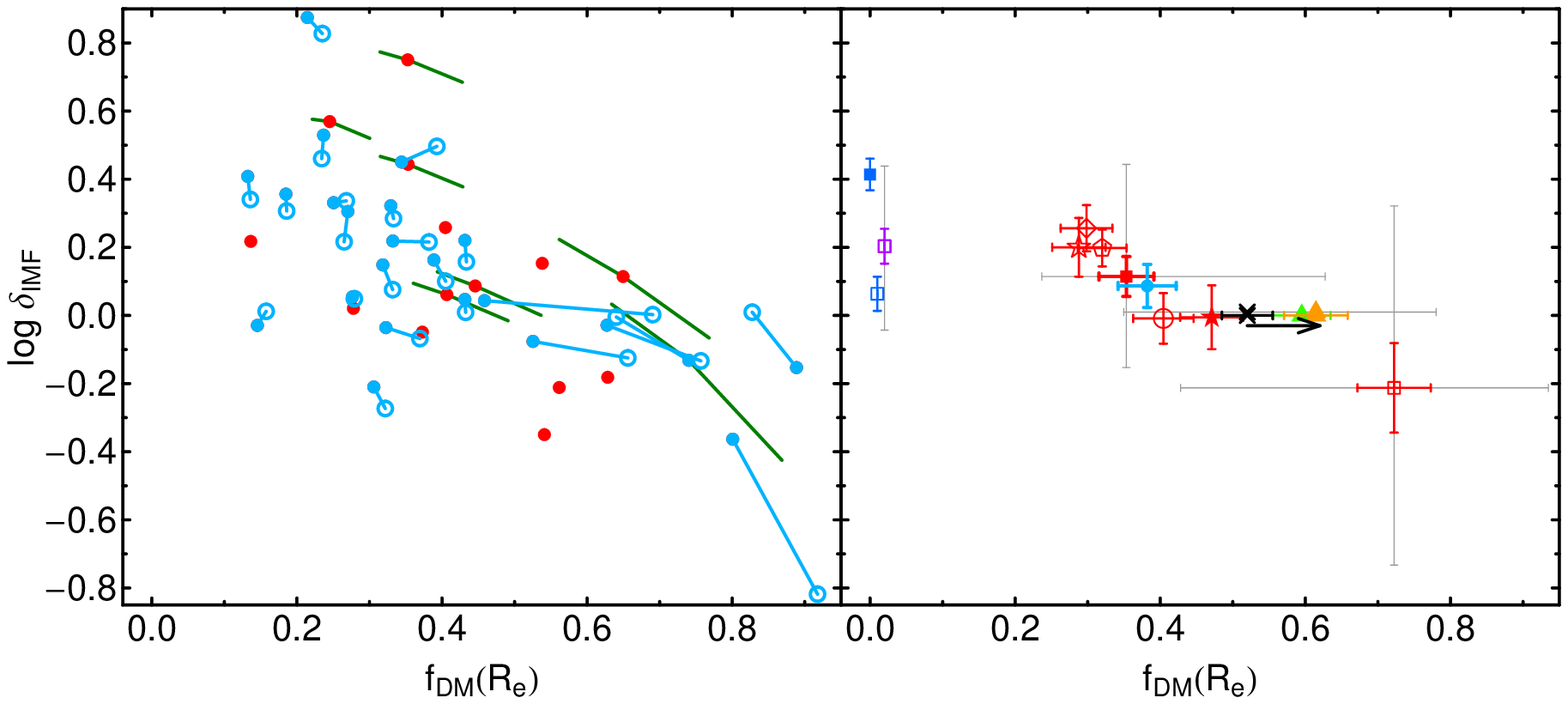, width=0.77\textwidth}
\psfig{file=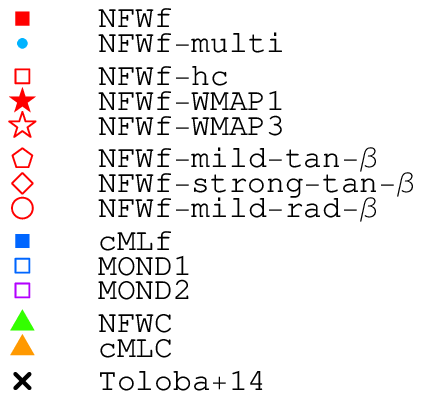, width=0.22\textwidth, bb=60 -20 200 100}
\caption{IMF normalization, \dimf , vs. effective ($<1 \Re$) DM
fraction, \fdm . Left. Results of the \NFWf\ model for individual
galaxies. For systems with no S\'ersic index available from
\citet{Janz+14}, we plot the range of values found if $n$ is
varied from $0.5$ to $2$ (dark--green curves). Cyan symbols
correspond to galaxies with either a single (dots) or
double--S\'ersic (open circles) fit available. Different cyan
symbols are connected by cyan lines to highlight the effect of
changing the parametrization of the light profile on our results.
Red points are for galaxies for which only single--fit parameters
are available. Right. Medians and standard errors of the median
are plotted for different models (see Table~\ref{tab:tab1}):
\NFWf\ (filled red square), \NFWfmulti\ (cyan dot), \NFWfhc\ (open
red square), \NFWfWMAPone\ (filled red star), \NFWfWMAPthree\
(open red star), \NFWfmtani\ (open red pentagon), \NFWfstani\
(open red diamond), \NFWfmrani\ (open red circle), \cMLf\ (filled
blue square), \MONDone\ (open blue square), \MONDtwo\ (open violet
square), \NFWC\ (green triangle) and \cMLC\ (orange triangle). For
few representative models the 16-84th quantiles of the
distributions are also shown as gray error bars. The median \fdm\
and the standard errors of the median from T+14 are shown as a
black cross and error bars, and the change after uniforming the
definition for \Mdyn\ and \mst\ to ours is outlined with a black
horizontal arrow. Legend for the symbols plotted in the right
panel is also shown.}\label{fig:fdm_vs_dIMF}
\end{figure*}

To further explore the effect of mass modelling, as well as the
impact of adopting alternative theories of gravity on our results,
we also consider the following models.
\begin{itemize}
\item \cMLf. In contrast to the DM haloes predicted by N-body simulations, a different class of models
relies on the assumption that total mass follows the light
distribution, i.e. constant \ML\ models with $\rho_{\rm M/L}=
\Ytot j_{*}$. Thus, we adopt a no-DM model with constant \ML\
profile, defined to have a total mass distribution $\rho_{no DM}=
\Yst j_{*}$, with \Yst\ as the only free fitting parameter.
\item \MOND . A modified Newtonian gravitational acceleration
model, in the regime of low acceleration, according to the MOND
theory (\citealt{Milgrom83b}; \citealt{Begeman+91}), has become an
alternative theory to reproduce galactic dynamics without DM. The
acceleration as a function of the radius $r$, $g(r)$, is given by
$g(r) \mu \left(x \right) = \gN(r)$,  where $x=g(r)/\a0$, \a0 is
the MOND acceleration constant (which sets the transition from the
Newtonian to the low acceleration regime), \gN\ is the Newtonian
acceleration, and $\mu(x)$ is an empirical function interpolating
between the two regimes. We adopt the following expressions: a)
$\mu_{1}(x)= x/(1+x)$ (\MONDone, \citealt{Famaey_Binney05};
\citealt{Angus08}) and b) $\mu_{2}(x)= x/\sqrt{1+x^{2}}$, which
has been the first one successfully tested with observations
(\MONDtwo, \citealt{Sanders_McGaugh02}). A constant \ML\ profile
with a free \Yst\ is adopted for the total mass distribution (see
\citealt{Tortora+14_MOND} for further details).
\end{itemize}

To complete our large model portfolio, we also adopt two models
with $\Yst = \Yssp$, with a fixed Chabrier IMF normalization:
\begin{itemize}
\item \NFWC . A NFW model, adopting the same \Mvir--\mst\ relation used for \NFWf, but dismissing the $\cvir-\Mvir$ relation, and
leaving \cvir\ as a free parameter.
\item \cMLC . A total mass distribution with radially constant \ML , but
setting $\Yst = \Yssp$. The total \ML\ is left free to vary. This
model is characterized by a radially constant DM fraction.
\end{itemize}

The final products of our analysis are the SSP \ML, \Yssp, the
dynamically-determined stellar and total \ML's, \Yst\ and \Ydyn,
the inferred mismatch parameter \dimf\ (for models with free IMF
normalization), defined as $\dimf\ \equiv \Yst/\Yssp$, and
effective DM fraction, $\fdm \equiv 1-\Yst/\Ydyn$.

\begin{table*}
\centering \caption{Mass models adopted in this study. Column~1
reports the label of each model. Columns~2, ~3 and~4 list the main
model ingredients, while columns~5 and~6 give the corresponding
IMF normalization, \dimf, and DM content, \fdm, estimates from our
dynamical analysis. Median, 16-84th percentiles of the sample
distribution and the standard error of the median are
shown.}\label{tab:tab1}
 \resizebox{.7\textwidth}{!}{\begin{tabular}{lccccc} \hline \hline
ID & Model & \Mvir\ - \cvir\ & IMF &  \multicolumn{2}{c}{Results}\\
(1)   &  (2)     &  (3)   &  (4) &   (5) & (6) \\
   &       &   &   &    $\log \dimf$  &   \fdm\  \\
\hline
\NFWf & NFW+S\'ersic & M+10 - WMAP5 & free & $0.11_{-0.27}^{+0.33}$ $\pm 0.06$ & $0.35_{-0.12}^{+0.27}$ $\pm 0.04$   \\
\NFWfmulti & NFW+2 S\'ersic & M+10 - WMAP5 & free & $0.09_{-0.22}^{+0.36}$ $\pm 0.06$ & $0.38_{-0.15}^{+0.27}$ $\pm 0.04$  \\
\\
\NFWfhc & NFW+S\'ersic & M+10 - $\cvir=20$ & free & $-0.21_{-0.52}^{+0.53}$ $\pm 0.13$ & $0.72_{-0.29}^{+0.21}$ $\pm 0.05$  \\
\NFWfWMAPone & NFW+S\'ersic  & M+10 - WMAP1 & free & $0_{-0.30}^{+0.37}$ $\pm 0.09$ & $0.47_{-0.16}^{+0.25}$ $\pm 0.04$   \\
\NFWfWMAPthree & NFW+S\'ersic  & M+10 - WMAP3 & free & $0.20_{-0.21}^{+0.22}$ $\pm 0.09$ & $0.29_{-0.09}^{+0.20}$ $\pm 0.04$  \\
\\
\NFWfmtani & NFW+S\'ersic+$\beta=-0.4$  & M+10 - WMAP5 & free & $0.20_{-0.19}^{+0.31}$ $\pm 0.05$ & $0.32_{-0.10}^{+0.24}$ $\pm 0.03$  \\
\NFWfstani & NFW+S\'ersic+$\beta=-1$  & M+10 - WMAP5 & free & $0.26_{-0.21}^{+0.22}$ $\pm 0.07$ & $0.30_{-0.10}^{+0.20}$ $\pm 0.04$  \\
\NFWfmrani & NFW+S\'ersic+$\beta=0.4$  & M+10 - WMAP5 & free & $0.0_{-0.28}^{+0.37}$ $\pm 0.07$ & $0.40_{-0.14}^{+0.25}$ $\pm 0.04$  \\
\\
\cMLf & constant \ML\ & - & free  & $0.41_{-0.23}^{+0.24}$ $\pm 0.05$ & 0    \\
\MONDone & $\mu_{1}$+constant \ML\ & - & free  & $0.06_{-0.22}^{+0.25}$ $\pm 0.05$ & 0 \\
\MONDtwo & $\mu_{2}$+constant \ML\ & - & free  & $0.20_{-0.24}^{+0.24}$ $\pm 0.05$ & 0 \\
\\
\NFWC & NFW+S\'ersic & M+10 - \cvir\ free & Chabrier  & 0 & $0.60_{-0.24}^{+0.16}$ $\pm 0.04$  \\
\cMLC & constant \ML\ & - & Chabrier   & 0 & $0.61_{-0.26}^{+0.17}$ $\pm 0.04$  \\
\hline
\end{tabular}}
\end{table*}

\section{IMF mismatch and DM fractions}\label{sec:results}

\begin{figure*}
\centering \psfig{file=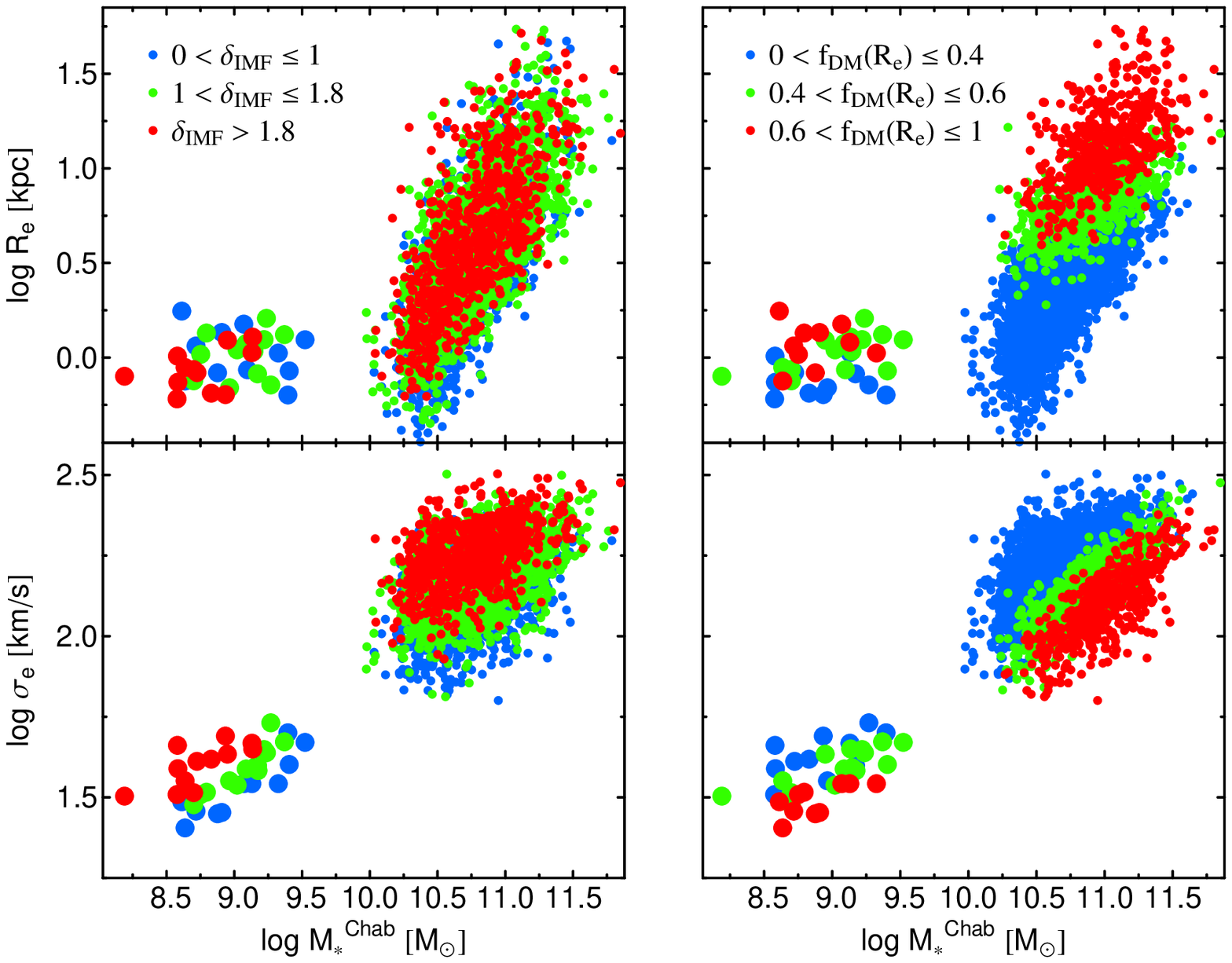, width=0.8\textwidth}
\caption{Effective radius (top-) and galaxy velocity dispersion
(bottom-panels) versus $\mst^{\rm Chab}$. Different points are
color-coded according to \dimf\ (left panels) and \fdm\ (right
panels), respectively (see labels in the top panels). The model
\NFWf\ is adopted. Large and small dots correspond to SMAKCED dEs
and bright SPIDER ETGs, respectively. Structural parameters for
the SPIDER ETGs have been measured in the K band, roughly matching
the H-band photometry of SMAKCED dEs.}\label{fig:size_sigma_mstar}
\end{figure*}

In Figure~\ref{fig:fdm_vs_dIMF} we plot the results of our
analysis. In the left panel, we plot \dimf\ vs. \fdm\ for our
reference \NFWf\ model (cyan symbols), while the median \dimf,
computed for the whole sample of dEs, versus the median \fdm, are
shown in the right panel, for all models listed in
Table~\ref{tab:tab1} (error bars show the 16-84th percentile
scatter in the data). For the ``standard'' \NFWf, in two cases
(i.e. VCC 0009 and VCC 1355) the model fails to fit the data. Only
50\% of the sample is fitted by the ``high-concentration'' model,
\NFWfhc. For \NFWfWMAPone , VCC 0009 is the only galaxy for which
the model fails, while for \NFWfWMAPthree\ the model fails for VCC
0009, VCC 0170, VCC 0308 and VCC 1355. The ``fixed IMF'' models
(\NFWC\ and \cMLC) only fail for VCC 1910. On the other hand,
\MOND\ and \cMLf\ models allow all galaxies to be fitted. The
two-component (or Chabrier IMF-fixed) models fail to reproduce the
data as the mass from the assumed DM model (or from the Chabrier
IMF-based model) is larger than the total mass allowed by the
observed \sige. This is not contemplated in the \cMLf\ and MONDian
models, for which it is always possible to find a \Yst\
reproducing the data.

In the next sections we will discuss the \dimf\ and \fdm\
estimates for each galaxy in the sample and then we will study the
impact of mass model comparing the median and the standard error
of the median\footnote{In statistics, the standard error of the
mean tells how accurate the estimate of the mean is likely to be,
and it is different from the standard deviation of a set of data.
A similar definition can be made for the median, and in
particular, it can be shown that the standard error of the median
is obtained by multiplying the standard error of the mean by the
factor 1.253 (\citealt{Harding+14_st_err}).} calculated over the
sample distribution of the models in \Tab\ref{tab:tab1}. We also
determine the 16-84th percentiles of the distributions.

\subsection{NFW and systematics in light profile}\label{subsec:IMF_DM}

Our reference \NFWf\ model produces, on average, $\dimf \sim 1.3$
($\log$~\dimf$\sim 0.1$ dex) and $\fdm \sim 0.35$.  The large
scatter in \dimf\ (see points in the left panel and gray scatter
bar in the right panel of Figure~\ref{fig:fdm_vs_dIMF}) is due to
the fact that while most galaxies turn out to have an IMF
normalization consistent with a MW-like distribution
($\log$~\dimf$\sim 0$), a significant fraction of them ($\sim
30\%$) have a super-Salpeter IMF normalization (i.e. $\dimf >
1.8$). DM fractions are, on average, in the range $\sim 0.2$ to
$0.6$, consistent with independent estimates for ``normal'' (as
opposite to dwarf) ETGs (see below). The standard error of the
median is $\sim 0.04$ for \fdm\ and $\sim 0.05$ dex for \dimf .

Adopting $10\%$ uncertainties on \Re\ and n, and propagating the
uncertainties on age and metallicity from Table~5 in T+14, we find
$1\sigma$ average errors on \dimf\ and \fdm\ of $\sim 0.15$ dex
and $\sim 0.05$, respectively. Taking the uncertainties into
account, we find that some dEs have \dimf\ inconsistent (heavier
than) a Chabrier IMF normalization. In particular, at the
$3\sigma$ level, the galaxies VCC0397, VCC0750 and VCC1684 have
$\dimf> 1.8$, while VCC0523, VCC0781, VCC1122 and VCC1528 have
$\dimf> 1$ (but $< 1.8$). Assuming $30\%$ (rather than $10\%$)
uncertainties on both \Re\ and n, the error on \dimf\ is almost
unchanged, while for \fdm\ is of $\sim 0.1$ and we find that, at
$3\sigma$, VCC0750 and VCC1684 still have $\dimf
> 1.8$, while VCC0397 and VCC0781 have $1 < \dimf < 1.8$. Note
that to these larger \dimf\ correspond smaller DM fractions, with
$\fdm \sim 0.2$--$0.3$. Thus, while for most galaxies our results
are consistent with a MW-like normalization, our analysis suggests
that for at least a few dEs, the \ML\ detected from previous
studies (e.g. T+14) might be partly accounted for by heavier IMF
normalization, rather than large DM fractions in the galaxy
central regions.

Note that a \dimf\ larger than one can be due to either a
bottom-heavy (due to a larger fraction of dwarf relative to giant
stars) or a top-heavy IMF (because of the large fraction of
stellar remnants from evolved massive stars). This degeneracy has
been broken in ETGs, by studying gravity-sensitive features in the
integrated light of galaxies. Such spectroscopic approach allows
one to constrain the mass fraction of dwarf-to-giant stars in the
IMF, rather than its overall normalization, in contrast to
dynamical and lensing methods (\citealt{Conroy_vanDokkum12b};
\citealt{Spiniello+12}; \citealt{LaBarbera+13_SPIDERVIII_IMF}).
Indications of top-heavy IMFs are found from a) galaxy number
counts (\citealt{Baugh+05}), b) in ultra compact dwarfs, based on
the large fraction of low-mass X-ray binaries found
(\citealt{Dabringhausen+12}), c) in the MW center (e.g.,
\citealt{Bartko+10}) or d) in Galactic globular clusters (e.g.,
\citealt{Prantzos_Charbonnel06}). In addition, galaxy formation
models reproduce the observed Intra-Cluster medium if a top-heavy
IMF is adopted (\citealt{Nagashima+05}). Studies of
gravity-sensitive features in dEs are still missing, thus, we
cannot exclude a top-heavy IMF in these systems.

As mentioned in \Sec\ref{sec:data}, S\'ersic indices are measured
only for 30, out of 39, dEs analyzed in this work. For systems
with no available n, we have adopted $n=1$. To test the effect of
this assumption, we have varied the S\'ersic $n$ from $0.5$ to
$2$, which is a ``conservative'' range, encompassing the observed
values, for the SMAKCED sample of dEs. For the 7 (out of 9)
systems where the \NFWf\ fits do not fail, the impact of changing
the $n$ is shown in the left panel of \Fig\ref{fig:fdm_vs_dIMF}
(see dark--green lines), which show how the \dimf\ and \fdm\
values change when the $n$ is varied. Larger S\'ersic indices
correspond to smaller \dimf\ and larger \fdm. We find, on average,
a mild variation of $\sim \pm 0.05$~dex on both \dimf\ and \fdm,
with negligible impact on median values for the whole sample. We
have further tested the effect of the $n=1$ assumption on the
galaxies with a measured S\'ersic index, by varying it to $n=1$.
On average, the variation is $0.03$ for \fdm\ and $-0.02$ dex for
\dimf .

We have also tested the impact of different parametrization of the
light on our dynamical estimates. The model \NFWfmulti\ replaces
the single S\'ersic profile with the double S\'ersic
parametrization for those systems which are better fitted by the
latter model. The results over the whole sample for both \NFWf\
and \NFWfmulti\ models are plotted in the left panel of
\Fig\ref{fig:fdm_vs_dIMF}, to highlight the effect of a refined
description of the light distribution in our dynamical analysis.
The galaxies with the largest \fdm\ are the most affected.
However, the medians computed over the whole sample are almost
unchanged with respect to the reference \NFWf\ (see right panel of
\Fig\ref{fig:fdm_vs_dIMF}).

As we have mentioned in \Sec\ref{subsec:data}, galactic nuclei, if
present, are excluded in the fit of the light distribution of the
SMACKED dEs. It is important to quantify the impact on our results
if these nuclei are included in the analysis. Because it is not
trivial to extract the amount of light due to the nuclei from
T+14, we rely on estimates from independent literature. 9 out of
the 39 galaxies are shared with \cite{Paudel+11}, who has provided
estimates of nuclear fluxes, $f_{\rm nuc}$, at $r < 2''$ and total
ones, $f_{\rm tot}$ (their Table 1), finding that the nuclei
account for $< 2\%$ of the total flux. Assuming that this limit
can be extended to the whole SMAKCED sample, we have included in
our modelling procedure a constant mass distribution with flux
$f_{\rm nuc}$ at $r < 2''$. We find that the impact on our results
is negligible, since the \dimf\ gets smaller by $0.02$ dex, while
the \fdm\ are left unchanged. The \dimf\ would get smaller by more
than $0.1$ dex for nuclei which account for more than $10\%$ of
the total flux. However, such prominent nuclei are not observed.

Finally, correcting for not detected emission lines would make the
ages and stellar masses smaller, getting \fdm\ and \dimf\ larger.
To provide a quantification of this effect we have augmented the
measured $H_{\rm \beta}$ index by its error (which we use to
estimate the impact of emission), and matched it with
\cite{Vazdekis+12} model predictions, deriving younger stellar
populations. We have also taken into account metallicity change
due to age variation, using the model predictions on the
$[Mg/Fe]'$--$H_{\rm \beta}$ plane. On average, stellar masses and
\dimf\ get smaller and larger by $\sim 0.05$ dex, respectively,
and \fdm\ is larger by $\sim 0.07$.

\subsection{DM mass model degeneracy}\label{subsec:DM_degeneracy}

We start discussing the impact of different \cvir--\Mvir\
relations. We notice that for ``high-concentration'' NFW models,
\NFWfhc, the best-fitting \Yst\ is significantly lower than that
for a Chabrier IMF (i.e. \dimf $< 1$). As already discussed, a
value as high as $\cvir = 20$ is able to describe only $\sim 50\%$
of the SMAKCED dEs. These models predict too much mass in the
center than what is allowed by the measured \sige . Together with
the fact that the Chabrier IMF gives the minimum normalization
with respect to either top- or bottom-heavier distributions (when
other relevant stellar population parameters, such as age and
metallicity, are fixed), this result suggests that high
concentration models are generally disfavoured for SMAKCED dEs.
Since only 50\% of the sample is fitted, the standard errors of
the median get higher, i.e. $\sim 0.05$ for \fdm\ and $\sim 0.13$
dex for \dimf . For what concerns different theoretical
predictions for the \cMvir\ correlation, we have analyzed the
impact of different cosmologies (WMAP1 vs. WMAP3 vs. WMAP5). We
find that within a WMAP1 cosmology, which predicts larger
concentrations with respect to WMAP5, \dimf\ gets smaller ($\sim
1$), and DM fraction larger than the WMAP5 case. In contrast, a
WMAP3 cosmology, because of smaller concentrations, gives larger
\dimf\ and smaller DM fractions (see \Fig\ref{fig:fdm_vs_dIMF}).

The right panel in  \Fig\ref{fig:fdm_vs_dIMF} illustrates the
effect of anisotropy on our sample results (see open red symbols),
with respect to our reference \NFWf\ model (filled red square).
For tangential anisotropy (e.g. \citealt{Geha+02}), we obtain
larger dynamical masses (see also~\citealt{SPIDER-VI}), larger
\dimf\ and smaller \fdm\ with respect to the \NFWf\ isotropic
case. In particular, for $\beta=-0.4$ ($\beta=-1$) \dimf\ gets
larger by $\sim 0.09$ ($\sim 0.15$) and \fdm\ is smaller by $\sim
0.03$ ($\sim 0.05$). For the sake of completeness, we also
consider here results when radial anisotropy is assumed. This
provides smaller \dimf\ (by $\sim 0.11$) and larger DM content (by
$\sim 0.05$). Overall, we conclude that the effect of anisotropy
is not negligible, especially for what concerns \dimf , while for
\fdm\ the results remain constrained within the typical error
budget of our models, even with the more ``extreme'' assumption
(e.g. $\beta=-1$).

The highest \dimf\ are obtained, as expected, by the model with no
DM, i.e. \cMLf\ (filled blue square and error bar in
Figure~\ref{fig:fdm_vs_dIMF}), where \dimf\ is significantly
larger than the \NFWf\ case ($\dimf > 1.5$), with $\sim 80\%$ of
galaxies nominally consistent with a super-Salpeter IMF
normalization. When assuming tangential anisotropy, all galaxies
in this model would turn out to have super-Salpeter normalization.

For what concerns models with modified gravity, median results are
shown in Figure~\ref{fig:fdm_vs_dIMF} as open blue (\MONDone) and
violet squares (\MONDtwo), respectively. The \dimf's differ by
$\sim 0.14$~dex between the two cases, bracketing results for the
\NFWf . However, within uncertainties the two models give
consistent results. Thus for dEs we confirm the same conclusions
as for more massive ETGs, i.e. that the MOND and DM frameworks are
equivalent to reproduce the dynamics of the central galaxy regions
(\citealt{Tortora+14_MOND}).

Finally, Figure~\ref{fig:fdm_vs_dIMF}  shows \fdm\ estimates
for the two models with fixed Chabrier IMF, i.e.~\NFWC\ and
\cMLC , respectively. As expected,
because of the intrinsically lower stellar \ML\ normalization,
the \fdm\ are much larger ($\gsim 0.4$, with an
average of $\sim 0.6$) than the \NFWf\ model.

\subsection{Comparison with literature}

As a comparison with independent results,
Figure~\ref{fig:fdm_vs_dIMF} also plots \fdm\ estimates from
Table~8 of T+14 (black cross and arrow in the right panel in \Fig
\ref{fig:fdm_vs_dIMF}). They adopt a Kroupa IMF to describe the
stellar component and use the virial relation $\Mdyn \propto K(n)
\sigma^{2} \Re$, with $K=3.63$, corresponding to a S\'ersic
profile with $n \sim 2$ (\citealt{Cappellari+06}). After
correcting their stellar masses to a Chabrier IMF, we obtain a
median \fdm\ value of $0.52_{-0.15}^{+0.16}$, which can be
compared with our results for the \cMLC\ model. A difference of
$\Delta \fdm \sim 0.1$ is found (see \Fig\ref{fig:fdm_vs_dIMF}).
This discrepancy is due to the fact that 1) T14 adopt a different
definition, with respect to our work, for the total and stellar
mass within 1 \Re, as dynamical masses are estimated within a
sphere with radius \Re\ (following the virial definition), while
stellar masses are calculated within a projected cylinder with
radius \Re; 2) the average S\'ersic index of the T14 sample is $n
\sim 1.5$, and not $n \sim 2$ (as they assume to compute the
$K(n)$). Note that the $K(n)$ tends to decrease with $n$
(\citealt{BCD02}). Correcting for these different assumptions,
\Mdyn\ become larger and stellar masses within 1 \Re\ are smaller,
making the median \fdm\ value larger and identical to our \cMLC\
estimate (this effect is outlined by the black horizontal arrow in
the right panel in \Fig\ref{fig:fdm_vs_dIMF}). This agreement is
expected if the same data and mass model are adopted.

We have also performed a comparison of our \cMLC\ \fdm\ with the
values derived by means of a complete Jeans dynamical modelling
(JAM) of the 2D kinematics in \cite{Rys+14}. After the
cross-matching, 6 galaxies are left. Looking at their face values,
\cite{Rys+14} \fdm\ are, on average, larger by $\sim 0.1$. These
JAM values look quite similar to the virial predictions, which
assume the same K-value of massive ETGs in \cite{Cappellari+06}.
If we normalize these mass definitions to our \cMLC\ model, the
discrepancy is even larger (by $\sim 0.2$). However, we find that
this discrepancy is possibly related to inhomogeneity between the
SMAKCED and \cite{Rys+14} data sets as differences in a) wavebands
used to calculate structural parameters and \ML\ (H-band in
SMAKCED vs. r-band in \citealt{Rys+14}), b) stellar mass
determinations, since absorption lines are fitted with
\cite{Vazdekis+12} models in SMAKCED survey and color-to-\Yst\
formula from \cite{Bell+03} is used in \cite{Rys+14}, and c)
estimated velocity dispersions. A complete understanding of these
sources of systematics is beyond the scope of the present work.

\subsection{Comparison with massive ETGs}
\label{subsec:comp}

In this section, we discuss our results for dEs into  the broader
framework of the continuity of intrinsic properties of spheroidal
systems, comparing the finding for dEs with those for a local
($0.05<z<0.095$), complete, sample of $\sim 4300$ giant ETGs drawn
from the SPIDER survey (see \citealt{SPIDER-I} and
\citealt{SPIDER-VI} for further details about sample selection).
The SPIDER dataset includes stellar masses derived from the fit of
stellar population synthesis (SPS) models to
optical$+$near-infrared photometry\footnote{ Note that these
stellar masses are consistent with ones obtained adopting
absorption lines (\citealt{SPIDER-V}).} (\citealt{SPIDER-V}),
galaxy structural parameters (effective radius \Re\ and S\'ersic
index $n$; using 2DPHOT, \citealt{LaBarbera_08_2DPHOT}),
homogeneously derived from $g$ through $K$ wavebands, and SDSS
central-aperture velocity dispersions, $\sigAp$. SPIDER ETGs are
defined as luminous bulge dominated systems, featuring passive
spectra in the central SDSS fiber aperture (\citealt{SPIDER-I}).
The dynamical analysis, presented in our previous papers for
SPIDER ETGs (\citealt{TRN13_SPIDER_IMF};
\citealt{Tortora+14_DMslope}), is similar to that carried out for
dEs in the present work. In particular, we have derived DM content
and IMF normalization for SPIDER ETGs using the \NFWf\ profile,
i.e. assuming the NFW+S\'ersic profile, with the same
\cvir--\Mvir\ and \mst--\Mvir\ relations as for dEs.

\begin{figure*}
\centering \psfig{file=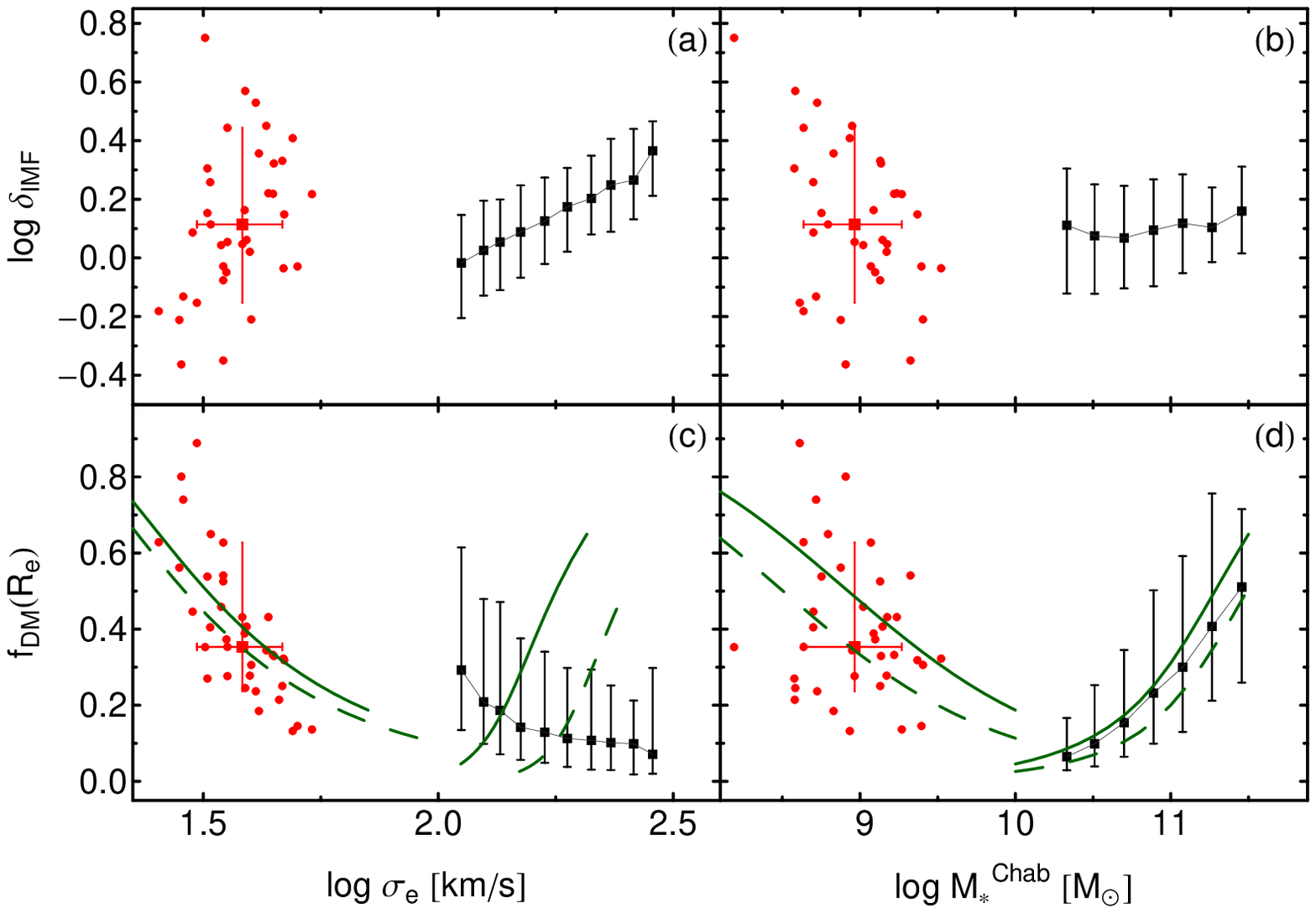, width=0.75\textwidth}
\caption{The \dimf\ (top) and \fdm\ (bottom) are plotted as a
function of galaxy velocity dispersion, \sige\ (left) and
Chabrier-based stellar mass, $\mst^{\rm Chab}$ (right). Red and
black symbols are for dE and SPIDER ETGs, respectively. For the
SPIDER sample, only median values with 16-84th quantiles are
shown, as black squares with error bars. The red squares and error
bars plot median values and 1~$\sigma$ scatter for the sample of
dEs (plotted individually as red dots). Green lines are
toy-models, obtained assuming a \NFWf\ model, and mean size- and
n-$\mst^{\rm Chab}$ relations derived, separately, from the two
samples. Solid and dashed lines refer to $\dimf = 1$ (MW-like) and
$=1.8$ (Salpeter normalization),
respectively.}\label{fig:dimf_fdm_mstar}
\end{figure*}

The results for dE and luminous ETGs are compared in
Figure~\ref{fig:size_sigma_mstar}, where \Re\ and \sige\ are
plotted as a function of \mst\ (see also T+14), and colour-coded
in terms of \dimf\ and \fdm\ (left- and right-panels),
respectively. The figure shows the well-known shallower \Re--\mst\
and \sige --\mst\ relations for dEs, with respect to the relations
for luminous ETGs (\citealt{Matkovic_Guzman05}; \citealt{Woo+08};
\citealt{Toloba+12_II}). Note that dEs have an almost constant
\Re\ with respect to \mst, with no dependence of \dimf\ on \Re\ at
fixed stellar mass (see big dots with different colors in the
top--left panel of Figure~\ref{fig:size_sigma_mstar}). On the
contrary, in the \sige--\mst\ plot (bottom--left panel), larger
\dimf\ correspond to larger \sige . Notice that this is expected
in our one-parameter \NFWf\ models, as dEs have almost constant
\Re , and at fixed \mst\ the only way to match a higher \sige\ is
thus to have a larger (smaller) \dimf\ (\fdm, see bottom--right
panel). For what concerns the behaviour with \fdm\ (right panels
of the Figure), we also notice that both dEs and luminous ETGs
have effective DM fractions driven by \Re, since bigger galaxies
have larger \fdm\  (see also \citealt{NRT10};
\citealt{SPIDER-VI}).

The trends of \dimf\ and \fdm\ with \mst\ and \sige\ are plotted
more explicitly for dEs and luminous ETGs, in
\Fig\ref{fig:dimf_fdm_mstar} (red and black symbols,
respectively). The \dimf\ (\fdm) appears to increase (decrease)
with \sige\ (panels (a) and (c)) consistent with what seen in
\Fig\ref{fig:size_sigma_mstar}, while no significant correlation
of both \fdm\ and \dimf\ is seen with $\mst^{Chab}$ (panels (b)
and (d)). However, we remind the reader that although
representative of the population of dEs, SMAKCED sample is
certainly incomplete with respect to \mst\ and \sige . Further
analysis, based on spatially extended kinematical data (i.e. 2D
spectroscopy) and complete galaxy samples, are required to
pinpoint the intrinsic correlations of DM and IMF normalization
with galaxy parameters in dEs.

For what concerns median values of \dimf\ and \fdm\ (red dots and
errorbars in \Fig\ref{fig:dimf_fdm_mstar}), we see that in the
\dimf--\mst\ diagram (panel (b)), the median \dimf\ for dEs is
consistent, overall,  with that for ETGs. This is due to the fact
that while most dEs have \dimf\ consistent with a MW-like
normalization, a significant fraction of them exhibit a Salpeter
or super-Salpeter normalization (\Sec\ref{sec:data}). In fact,
when looking at the \dimf--\sige\ plot (panel (a)), we see that
dEs tend to have, on average, slightly higher IMF normalization
than the lowest \sige\ ETGs (whose normalization is fully
consistent with the MW-like distribution). Interestingly, the
trend of \fdm\ with \mst\ (panel (d)) points to a double-value
behaviour of DM content with stellar mass in ETGs, with larger
\fdm\ in most massive ETGs ($\mst \gsim 10^{11}\, \rm \Msun$) and
dEs ($\mst \sim 10^{9}\, \rm \Msun$), and a minimum at $\mst \sim
10^{10}\, \rm \Msun$.

To have some further hints of the \fdm\ trends, we have
constructed toy mass models, based on our reference \NFWf\ model
by computing stellar mass profiles according to either a Chabrier
or a Salpeter IMF, and adopting the mean size- and $n$-mass
relations of dEs and massive ETGs, respectively. Results for the
\fdm\ trends are shown in the bottom panels of
\Fig\ref{fig:dimf_fdm_mstar}. At fixed IMF, such toy-models
predict a double-value trend for the \fdm\ as a function of both
\sige\ and $\mst^{\rm Chab}$ (panel c and d). In the $\fdm-\mst$
diagram (panel d), this trend matches quite well the observations
for both dEs and ETGs. In contrast, for the $\fdm-\sige$ diagram,
the toy-models give a good match to dEs (although some overall
variation of \dimf\ is required to exactly match the trend, as
also seen in panel a), while there is a clear mismatch in the case
of massive ETGs, where toy models predict an increasing trend of
\fdm\, while the inferred \fdm\ tends to midly decrease with \sige
. As seen in panel a, this disagreement is due to the fact that
toy models assume a constant IMF normalization, while data imply a
strong trend of \dimf\ with \sige . In summary, our toy-models
also support a double-value behaviour in the \fdm--\mst\
correlation.

We have also verified that the double-value trend of \fdm\ does
not depend critically on our assumptions of a given \Mvir--\mst\
relation. In fact, a similar result is found when considering the
\cMLC\ model.

The U-shape behaviour of \fdm\ with \mst\ in early-type systems
can be understood as a result of different feedback mechanisms in
these systems at different mass scales. In lowest mass galaxies
(dEs), star-formation is likely inhibited by (e.g.) supernovae
feedback. This becomes less important at increasing galaxy mass
(\citealt{dek_birn06}; \citealt{Cattaneo+08}). However, at the
highest \mst, additional processes, such as dry merging, AGN
feedback or halo-mass quenching further inhibit gas cooling, and
decrease star-formation efficiency again (\citealt{dek_birn06}).
Hence, the lowest- and highest-mass galaxies are expected to have
the lowest star-formation efficiency, and thus, under the
assumption of a universal DM distribution, the highest DM content.
The \fdm\ trend in dEs and luminous ETGs adds up to other well
known correlations in ETGs, such as the trends of total \ML\ and
star formation efficiency (\citealt{Benson+00}, \citealt{MH02},
\citealt{vdB+07}; \citealt{CW09}; M+10), the U-shape of half-light
dynamical \ML\ (\citealt{Wolf+10}; \citealt{Toloba+11_I}),
$\mu_e-\Re$ (\citealt{Capaccioli+92a};
 \citealt{Kormendy+09}) and size-mass (\citealt{Shen+03};
 \citealt{HB09_curv}) relations, the trends of optical colour and metallicity gradients
(\citealt{Spolaor+10}; \citealt{Kuntschner+10};
\citealt{Tortora+10CG}; \citealt{Tortora+11MtoLgrad}), as well as
DM gradients (\citealt{Napolitano+05}) with galaxy mass.

\section{Discussions and Conclusions}\label{sec:conclusions}

In this work, we have performed an isotropic Jeans dynamical
analysis for 39 dwarf ellipticals in the Virgo cluster, from T+14.
For the first time, we have studied the IMF normalization and the
effective DM content using a suite of fixed DM profiles and the
stellar \ML\ as a free-fitting  parameter. We have also performed
the analysis with MOND, modified-gravity, models. The main results
are shown in Figure~\ref{fig:fdm_vs_dIMF} where we find that, on
average, using a NFW profile and standard \cvir-\Mvir\ relation
from N-body simulations (\NFWf\ model in Table~\ref{tab:tab1}),
dEs have\footnote{Median and 16-84th percentiles of the sample
distribution are shown, together with the standard error.} $\log
\dimf = 0.11_{-0.27}^{+0.33}$ with error $\sim 0.05$ dex (i.e.
consistent with a Chabrier-like IMF normalization) and $\fdm =
0.35_{-0.12}^{+0.27}$ with error $\sim 0.03$. A constant-\ML\
model, with no--DM content (the \cMLf\ model), maximizes the
stellar mass content (i.e. the \dimf ), pointing to super-Salpeter
IMF normalizations. In the MOND scenario, using two standard
interpolating functions (\MONDone, \citealt{Famaey_Binney05};
\MONDtwo, \citealt{Sanders_McGaugh02}) we find results which
encompass the NFW predictions, in agreement with results for
massive ETGs (\citealt{Tortora+14_MOND}). For completeness, we
have also analyzed the cases of a universal Chabrier IMF (\NFWC\
and \cMLC), which provide larger effective DM fractions when
compared to our reference model with free IMF normalization, \NFWf
. The derived DM fractions for the \NFWC\ model are fully
consistent with the estimates in \cite{Toloba+14_II}, if stellar
and dynamical masses are homogeneously defined. We have also
analyzed the impact of several further assumption, such as
light-profile parametrization, velocity dispersion anisotropy, and
assumptions on the \cMvir\ relation. In particular, at the mass
scale of dEs, our data seem to disfavour a \cvir--\Mvir\ relation
steeper than that from simulations, as it might be the case for
more massive halos (see~\citealt{Leier+12}), while if tangential
anisotropy is assumed (see e.g. \citealt{Geha+02}), we obtain
larger \dimf\ and smaller \fdm\ with respect to the reference
isotropic case (on the contrary, radial anisotropy produces larger
\fdm\ and smaller \dimf). Although most of dEs have \dimf\
consistent with a MW-like normalization, for the reference \NFWf\
model, we also find evidence that some dEs might have $\dimf > 1$
or $> 1.8$ at high statistical significance, i.e. an IMF which is
heavier that a Chabrier- or Salpeter-like distribution.

In Figures \ref{fig:size_sigma_mstar} and
\ref{fig:dimf_fdm_mstar}, we have compared results for dEs with
those for massive ETGs from the SPIDER sample (\citealt{SPIDER-I};
\citealt{SPIDER-VI, TRN13_SPIDER_IMF}). We find some hints that
\dimf\ might increase with \sige, as in  more massive ETGs
(\citealt{TRN13_SPIDER_IMF,Tortora+14_DMslope}). However,
spatially extended kinematical data (i.e. 2D spectroscopy) and
complete galaxy samples are required to confirm if this trend is
real, rather than due to sample incompleteness.  Moreover, we find
that, on average, dEs tend to have slightly higher IMF
normalization than ETGs at lowest \sige\ ($\sim 100$~$\rm km
s^{-1}$ , whose IMF normalization is fully consistent with a
MW-like distribution).

The trend of \fdm\ with \mst\ suggests a double-value behaviour,
with largest \fdm\ in most massive ETGs ($\mst \gsim 10^{11}\, \rm
\Msun$) and dE's ($\mst \sim 10^{9}\, \rm \Msun$), and a minimum
at $\mst \sim 2-3 \times 10^{10}\, \rm \Msun$.  These trends
mirror those of the dynamical \ML\ (\citealt{Wolf+10};
\citealt{Toloba+11_I}), and of total star formation efficiency
with respect to mass (\citealt{Benson+00}, \citealt{MH02},
\citealt{vdB+07}; \citealt{CW09}; M+10), which are the result of
the interplay among different physical processes, such as SN
feedback at lowest galaxy masses, and AGN feedback, galaxy merging
and halo mass heating in the most massive ETGs
(\citealt{Tortora+10CG}).

In this paper, we have performed a first attempt to constrain,
simultaneously, the IMF normalization and dark matter content of
low-mass (dwarf) early-type galaxies, finding for at least a few
systems that the \dimf\ is heavier than the MW-like value. Since
such a ``heavy'' \dimf\ could be due to either a bottom- or a top-
heavy distribution, a natural follow-up of the present work would
be  to study gravity-sensitive features in the integrated light of
galaxies, which have provided, so far, important constraints to
the IMF slope of massive ETGs (\citealt{Conroy_vanDokkum12b};
\citealt{Ferreras+13} \citealt{LaBarbera+13_SPIDERVIII_IMF};
\citealt{Spiniello+12, Spiniello+14}). In fact, a similar analysis
is currently lacking for dE's. In the future, it will be also
necessary to apply the present analysis to large and complete
samples including a variety of stellar systems, such as dwarf
ellipticals, dwarf spheroidals, ultra-compact dwarfs, and
late-type galaxies, to achieve a complete picture of how the dark
matter and stellar components have been assembled along the whole
galaxy mass sequence.


\section*{Acknowledgments}

We thank the referee for his/her detailed report which has
contributed to improve the manuscript. CT has received funding
from the European Union Seventh Framework Programme
(FP7/2007-2013) under grant agreement n. 267251 ``Astronomy
Fellowships in Italy'' (AstroFIt). FLB acknowledges support from
grant AYA2013-48226-C3-1-P from the Spanish Ministry of Economy
and Competitiveness (MINECO). NRN acknowledges support from
Regione Campania L. 5/2002. The authors thank R. Peletier for the
useful comments and suggestions to improve the paper and E. Toloba
to have provided further information about the galaxy dataset.


\bibliographystyle{mn2e}   



\end{document}